\documentclass{ws-ijmpa2}

\begin{document}

\title{VARIABLE STARS MAGNITUDES ESTIMATIONS EXPLOITING THE EYE PHYSIOLOGY}

\author{COSTANTINO SIGISMONDI}

\address{Department of Physics, University of Rome "La Sapienza" and ICRA, 
International Center for Relativistic Astrophysics, P.le A.Moro 2 00185 
Rome Italy}

\maketitle

\begin{abstract}
The physiology of the dark adaption process of the eye 
is revisited from an astronomical point of view.
 A new method for the magnitude estimation of a star is presented.
It is based upon the timing of the physiological cycle of the rhodopsin 
during the eye dark adaption process.
The limits of the application of the method are discussed.
This method is suitable for bright stars as Betelgeuse, Antares or Delta Scorpii or stars at the limiting magnitude observed with a telescope.
\end{abstract}

\section{Introduction: the observations of variable stars}

Professional astronomers and astronomical observatories have not enough time for following all the light curves of the variable stars and novae appearing in the sky.
Such work is carried out by amateurs astronomers with very good skills, and they are gathered in the AAVSO international organization.\footnote{see http://www.aavso.org}
They use mainly naked-eye observations, because of the relatively high cost of CCD devices for amateur equipments.
The method of magnitude estimation of the star magnitude presented here (section two) is useful for observations near the telescope limiting magnitude (section three). The source of error in this method is outlined in section four.
The efficiency of this method with the altitude is also taken into account (section five). 

\section{The response to the darkness of the eye as magnitude estimator}

The physiology of the eye dark adaption is composed by two mechanisms:
\begin{itemize}
\item{the mechanical one}
in which the pupil reaches its maximum diameter ($\sim 7~mm$)
\item{the chemical one}
where the rhodopsin and iodopsin (for the cones) are regenerated in the retinal receptors. The rods and the cones have different time scales of chemical regeneration (5 minutes the former and 30 minutes the latter). 
\end{itemize}
Moreover the sensitivity to small intensity of light is much better in rods with respect to the cones (that are sensitive to the colours). Therefore the combination of all those factors gives the curve describing the dark adaption versus time shown in the figure 1 (adapted from Lerman (1980)\cite{boo}).

\begin{figure}
\centerline{\psfig{file=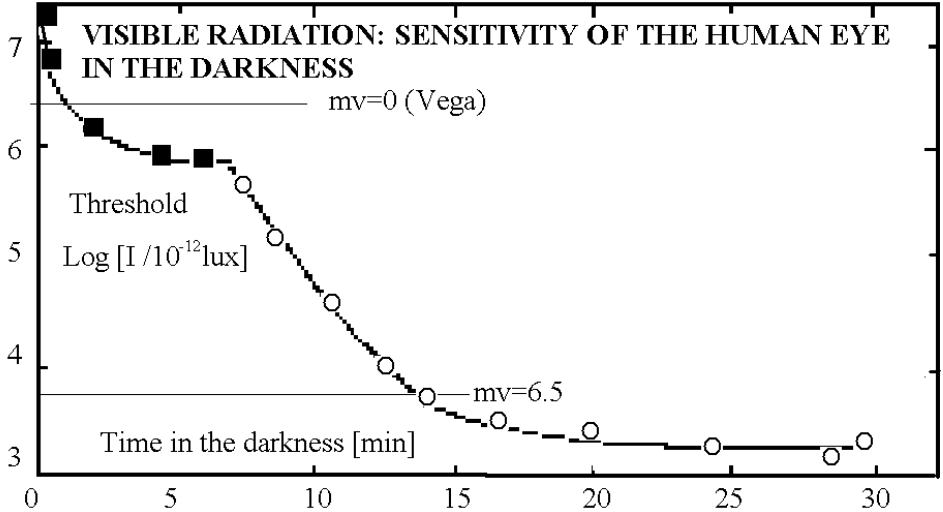,width=13cm}}
\vspace*{0pt}
\caption{The y axis is the logarithm of luminous intensity threshold visible to the naked eye measured in $\mu\mu$ lux. 
The x axis, in minutes, is the time during which the eye remains in the darkness.
Vega, a star of mv=0.03 has an intensity of $2.5\times 10^{-6}$ lux, i. e. 6.4 in the scale here represented. 
A star of mv=6.5, considered the fainter star detectable at the naked eye, corresponds to 3.8 in this scale.
The eye is sensitive to even fainter signals (3.3 in the y scale of this figure, i. e. 1.25 magnitudes fainter) 
but it is difficult to discern it as a star for its low signal-to-noise ratio. Figure adapted from Lerman (1980).}
\end{figure}

Such figure needs of a more extended comment with respect to the medical textbook, for an astronomical application.
The factor of 4 in logarithm of the intensity gained after 30 minutes of darkness corresponds to a gain of 10 magnitudes according to the Pogson's law. Under optimal optical conditions the naked-eye limiting magnitude is considered as $m_{lim}=6.5$ (see e.g. Jenniskens (1994)\cite{jen} for naked-eye meteor observations) for point-like sources. 
The initial point of such curve corresponds to $m_{v}\sim-3$. It corresponds indeed to daylight. A point-like object of $m_{v}\sim-3$ is indistinguishable from the background when the latter has a brightness of $5\frac{mag}{arcsec^2}$ (see appendix), say within $\sim10^o$ apart of the Sun with clear sky. We can also consider this value as the "bleaching (dazzling)" intensity of the light. That is a condition never occurring in the night because the home illumination does not reach that one of the sky near the Sun. The only one case of sudden decrease of luminosity is during the last stages of a total eclipse of the Sun, when the eye is not protected by appropriate filters.
So for astronomical use in figure 1 we can start directly by the knee occurring after 7 minutes of dark adaptions, with the reasonable assumption of do not start astronomical observations when bleached (dazzled). Moreover, when we start to observe the stars we are immediately able to distinguish the colours of the most bright, so the cones are already ready to detect and to analyse the star's light.

We can approximate the second part of the curve with the exponential law:
$m_{lim}(t)=m_0+3\cdot(1-\exp({-\frac{t}{\tau}}))~~~~~~~~~~~~~~~~~~(1)$ \linebreak

with $\tau=5~$minutes and $m_0$ the minimum magnitude visible at the beginning of the observation.
For small time intervals eq. (1) becomes $m_{lim}(\Delta t)=m_0+3\cdot\frac{\Delta t}{\tau}~~~~~~~~~~~~~~~~~~(2)$ \linebreak

It is to remark that the stars in the stellar field under examination should be all detected with undirect vision, i. e. not utilizing the fovea (the central region of the retina) where there are no rods. Starting the observations with a stellar field with known magnitudes, brighter than the variable star we measure the $\Delta t$ necessary to our eye, unadapted to the darkness, to see the variable star, once seen the comparison stars.
By the formula (1) we obtain the amount of $\Delta m$ with respect to the referring star $m_{var}=m_{ref}+\Delta m$.

\section{Applications}
 
This method allows to use the dark adaption time, commonly not exploited, as magnitude estimator. It is useful for objects with magnitude close to the limiting one, for which there are no fainter comparison stars visible. 
Analogously if the variable star is slightly brighter than the comparison star this method applies.
Using the telescope it is necessary to have an apparent field of view of almost $40^o$ in which the variable stars and the comparison stars are. 
Moreover a magnification of $100$x is  required for a better detection by the eye (Clark, 1997)\cite{sky}. Indeed for unaided eyes this method applies to some Mira-type variables around their maxima or other irregular bright stars like Antares, Betelgeuse and Delta Scorpii.\footnote{see http://www.icra.it/solar/fotometria-sigismondi.pdf} 
This method has been used also during the observations of V1483 AQL made by the author in 1999 five times with a 10 cm telescope at its limiting magnitude (12) from July 17 to July 22 while it was fading from 10.7 to 12 magnitude. The observations are registered in the AAVSO website with the observer's code SGQ.

\section{Experimental errors}

The own eye time response to the darkness is to be calibrated by timing when we see the stars with known magnitude in the selected field of view, after an exposition to the home-light. Once choosen the comparison stars, the source of error of this method is only in the timing of the appearance of the star under study, after having seen the reference ones.
So we deal with the determination of two time instants. Assuming an incertitude of 3 seconds in those determinations from eq. (2) we can have 
$\sigma(m)= 3\cdot\frac{\sigma(t)}{\tau}\sim0.03$ magnitudes (whit $\tau = 5$ minutes)~~~~~~~~~~~~~~~~~~(3)\linebreak

\section{The retinal response with the altitude}

The retina has the highest rate of respiration of any tissue in the body, i. e. maximum oxygen consumption per unit weight (Lerman, 1980). For naked-eye observations a site of 2700 meters of altitude is found equivalent to a 4600 meters-one (see e. g. Vanin, 1997\cite{van}). Even if the sky transparency is increased, the gain due to the quote is reduced by the decrease of the retinal functionality due to the reduction of available oxygen. Moreover 2700-2800 meters are the last altitudes convenient for naked-eye observations because of the scale height of oxygen disponibility in the terrestrial atmosphere following an exponential law (see e.g. Fermi (1936)\cite{fer}) $\rho_{O_2}=exp{(-\frac{h}{h_{scale}})}$ where $\rho_{O_2}$ is the oxygen density, $h$ is the quote and $h_{scale}\sim 1500$ m for the oxygen.

\section{Appendix}

For calculating the value of background luminosity when a point-like source of magnitude $m_v=-3$ is not visible because of dazzling conditions we should consider:
\begin{itemize}
\item{the pupil diameter in daylight: $\sim 3$ mm}
\item{the angular resolution of the eye (resolution power) according to Rayleigh criterion  in such illumination conditions: $\theta= \frac{120}{3}$arcsec.}
\item{the corresponding dimension of a pixel is 1600 arcse$c^2$.}
\item{a point-like object corresponds to a pixel}
\item{an object spanning 1 arcse$c^2$ is 1600 times fainter than the pixel dimension with constant surface brightness components}
\item{According to the Pogson's law the difference of magnitude between an arcse$c^2$ and 1600 of them is $\Delta(m)=2.5\cdot~log{1600}=8$}
\item{every arcse$c^2$ of background shines like a $5^{th}$ magnitude star.}
\end{itemize}

\end{document}